\documentclass[11pt]{article}

\usepackage{times}
\usepackage{latexsym}
\usepackage{hyperref}
\usepackage[numbers,sort&compress]{natbib}
\usepackage{geometry}
\usepackage[T1]{fontenc}
\usepackage[utf8]{inputenc}
\usepackage{microtype}
\usepackage{inconsolata}
\usepackage{graphicx}
\usepackage{color}
\usepackage{multirow}
\usepackage{makecell}
\usepackage{subfig}
\usepackage{soul}
\usepackage{mathtools}
\usepackage{bigstrut,multirow,rotating}
\usepackage{amsthm}
\usepackage{todonotes}
\usepackage{booktabs}

\newtheorem{definition}{Definition}

\title{Differentially Private Parameter-Efficient Fine-tuning for Large ASR Models}

\author{
  Hongbin Liu\thanks{Work done as a student researcher at Google.}$\:\;^{1,2}$, Lun Wang$^2$, Om Thakkar$^2$, Abhradeep Thakurta$^2$, Arun Narayanan$^2$\\
  $^1$Duke University, $^2$Google\\
  $^1$hongbin.liu@duke.edu, $^2$\{lunwang, omthkkr, athakurta, arunnt\}@google.com
}
\date{}

\begin{document}
\maketitle

\begin{abstract}
Large ASR models can inadvertently leak sensitive information, which can be mitigated by formal privacy measures like differential privacy (DP). However, traditional DP training is computationally expensive, and can hurt model performance. Our study explores DP parameter-efficient fine-tuning as a way to mitigate privacy risks with smaller computation and performance costs for ASR models. Through extensive experimentation and progressive optimization, we achieve 4.6\%/8.1\% word error rate on LibriSpeech clean/other test-sets, setting a new performance benchmark while maintaining $(10, 3.52\mathrm{e}{-6})$-DP in fine-tuning a large ASR model with over 600M parameters.
\end{abstract}
\section{Introduction}

\begin{figure}[!t]
    \centering
    \includegraphics[width=0.3\textwidth, trim = 6cm 0cm 6cm 0cm, clip, angle=-90]{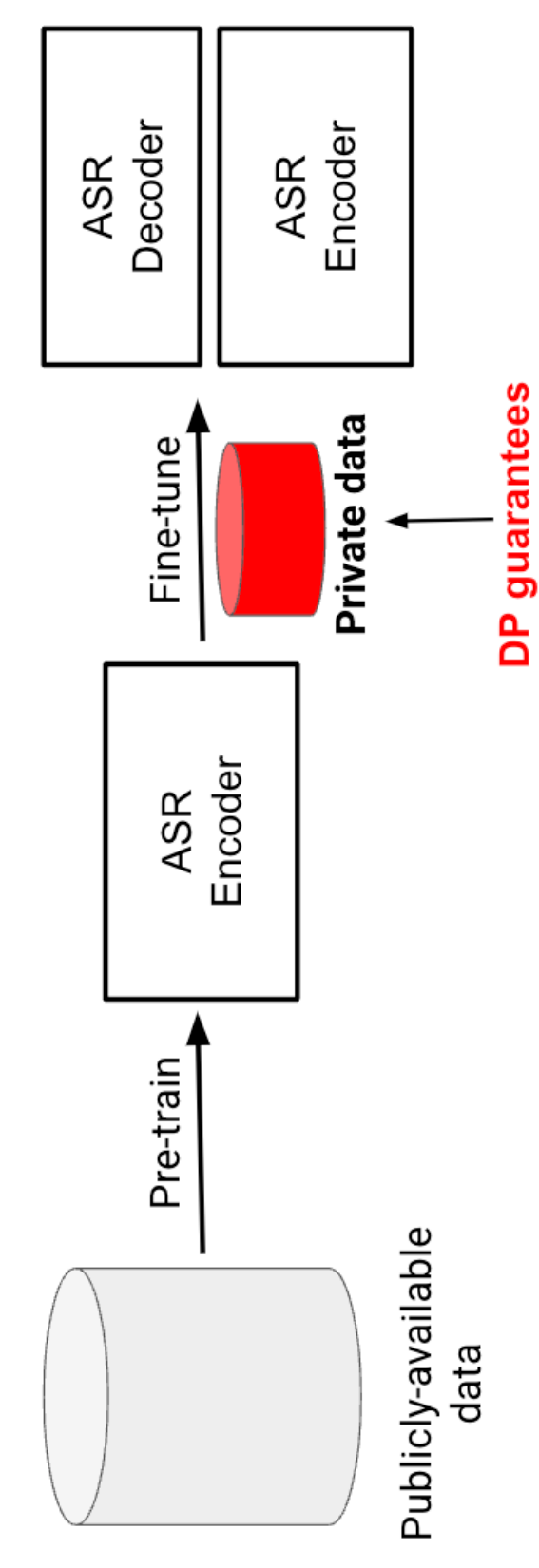}
    \caption{ASR model training workflow.
    }
    \label{fig:workflow}
\end{figure}

Recent years have witnessed remarkable advancement in Automatic Speech Recognition (ASR), with ASR models based on large speech foundation models, such as USM~\cite{zhang2023google} and Whisper~\cite{radford2023robust}, continuously pushing the boundaries of performance in speech recognition tasks. No matter integrated as a cascaded component or contributing to multimodal understanding, ASR also becomes crucial for processing spoken language in state-of-the-art large language models like Gemini 1.5~\cite{reid2024gemini} and GPT-4o~\cite{gpt4o}.
As shown in Figure~\ref{fig:workflow}, training cutting-edge ASR models typically begin with pre-training a foundation speech model, usually an ASR encoder, on vast amounts of publicly available data using self-supervision or weak supervision to learn general-purpose speech representations.
The pre-trained ASR encoder with a newly initialized ASR decoder are then fine-tuned on domain-specific datasets, that may contain sensitive or proprietary information, to adapt to specific downstream tasks.

Recent studies have revealed various privacy vulnerabilities~\cite{amid2022extracting,jagielski2022noise} in large ASR models, especially their potential to unintentionally memorize rare or unique samples from their fine-tuning data~\cite{wang2023unintended}.
The findings highlight the need for privacy-preserving training techniques, exemplified by differentially private stochastic gradient descent (DP-SGD)~\cite{abadi2016deep}, which has been the workhorse for DP training.
However, with increasing model sizes, DP-SGD often leads to decreased model performance~\cite{BST14,LLHIKLT22,de2022unlocking,liu2024pre}, and significantly increased computational costs~\cite{KM00TX21,yu2021not}, which can be a major obstacle as the cost of training large ASR models without privacy measures is already high.
A promising approach to balance privacy, utility and computation is to combine DP-SGD with parameter-efficient fine-tuning (PEFT).
For example, recent research~\cite{yu2021differentially,yu2021not,yu2021large,bu2022differentially} highlights the effectiveness of DP-PEFT in fine-tuning language models.

In this paper, we conduct \emph{the first comprehensive study of DP-PEFT in the context of ASR model fine-tuning}, and the findings are notably different from the language modeling domain.
We observe that \emph{DP-BitFit~\cite{zaken2021bitfit,bu2022differentially} emerges as the optimal method for both performance and computation costs}, while DP-LoRA~\cite{hu2022lora,yu2021differentially} and DP-Compacter~\cite{karimi2021compacter,yu2021differentially} are favored for performance and computation respectively in language models.

Our findings extend beyond a simple application of existing DP-PEFT methods to ASR models.
Instead, we conduct extensive, domain-specific optimizations to achieve these results.
First, we perform exhaustive ablation studies to refine existing DP-PEFT methods for ASR.
For example, we show that tuning specific sets of bias terms in DP-BitFit can significantly improve its performance. Similarly, we observe different strategies for parameter initialization and placement can result in improvements for DP-LoRA.
The latter investigation also leads us to propose a novel DP-LoRA variant, DP random projection (DP-RP), which halves trainable parameters while achieving comparable performance.
Second, our work is \emph{the first to showcase the potential of leveraging synthesized random audio samples to improve privacy-utility trade-offs for ASR DP fine-tuning}.
This innovative approach reduces the reliance on high-quality public data.
By incorporating this approach alongside other optimizations, we achieve state-of-the-art word error rates (WERs) on LibriSpeech test-sets with $(10, 3.52\mathrm{e}{-6})$-DP, establishing the efficacy of DP-PEFT for practical, privacy-preserving fine-tuning of large ASR models.

\section{Background and Related Work}

\subsection{Differential Privacy}
Differential privacy~\cite{dwork2006calibrating} is a rigorous mathematical framework for quantifying the privacy loss of an algorithm by comparing its output distributions on adjacent datasets, which differ by a single data point.
An algorithm is considered differentially private if its output distributions on adjacent datasets are statistically close.

\begin{definition}[Differential privacy~\cite{dwork2006calibrating}] 
A randomized function $\mathcal{F}: \mathcal{D}\rightarrow \mathcal{R}$  satisfies $(\varepsilon,\delta)$-DP if, for any two adjacent datasets $D$, $D^{\prime} \in \mathcal{D}$ and for any subset $S \subseteq \mathcal{R}$, it holds that:
\begin{align}
\text{Pr}[\mathcal{F}(D) \in S] \leq  e^\varepsilon \text{Pr}[\mathcal{F}(D^{\prime}) \in S] + \delta.
\end{align}
\end{definition}
In deep learning, $\mathcal{F}$ is the training algorithm such as SGD.
The parameter $\varepsilon$ is referred to as ``privacy budget'' and a smaller $\varepsilon$ means a stronger privacy guarantee.
DP-SGD is the primary method for differentially private deep learning.
It involves clipping and adding Gaussian noise to the gradients of each training example before updating the model parameters using any deep learning optimizer (we use Adam~\cite{kingma2014adam} in this paper).

\subsection{Parameter-Efficient Fine-Tuning}
\begin{figure}[!t]
    \centering
    \includegraphics[width=0.8\textwidth]{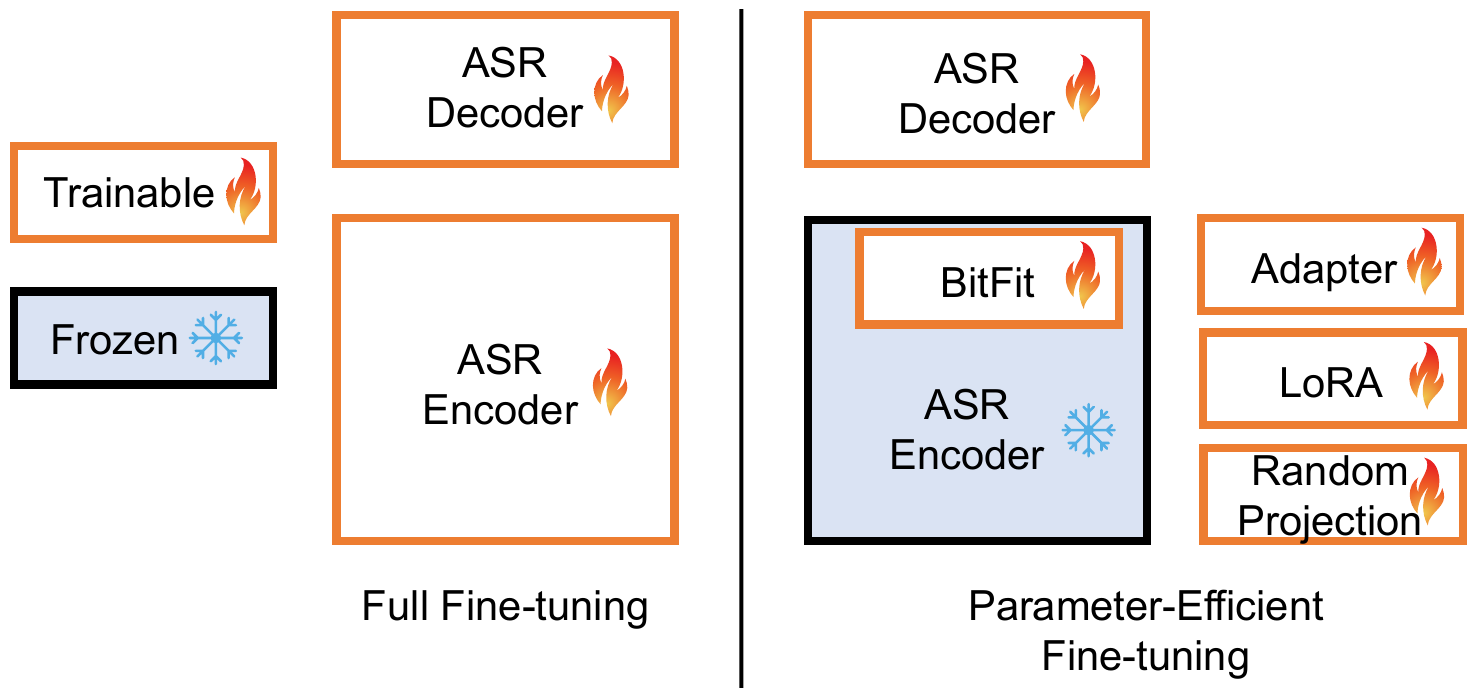}
    \caption{PEFT methods in ASR models.}
    \label{fig:FT_vs_PEFT}
\end{figure}

Parameter-efficient fine-tuning (PEFT)~\cite{houlsby2019parameter,hu2022lora,zaken2021bitfit,li2023modular,chen2023efficient} is a class of fine-tuning techniques that updates only a small number of either newly added, or existing parameters of a pre-trained foundation model.  
The first PEFT is Adapter~\cite{houlsby2019parameter}, which adds two low-rank projection matrices and one activation layer before the layer norm and after the feed-forward layer in each Transformer block~\cite{vaswani2017attention}, and only fine-tunes the added parameters.
LoRA~\cite{hu2022lora} adds and fine-tunes two low-rank projection matrices parallel to feed-forward layers.
BitFit~\cite{zaken2021bitfit} only trains the bias terms in a model.
Figure~\ref{fig:FT_vs_PEFT} illustrates these PEFT methods in the context of ASR models.

DP-PEFT methods, which optimize in lower-dimensional spaces, have been extensively studied due to potential benefits for DP~\cite{BST14,li2022when}. 
Notable work includes~\cite{yu2021large}'s reparameterized gradient perturbation,~\cite{yu2021differentially}’s DP-Adapter, DP-LoRA and DP-Compactor, and~\cite{bu2022differentially}'s DP-BitFit.

\section{Improving DP-PEFT for ASR}

In this section, we conduct extensive experiments to compare and refine DP-PEFT methods for ASR.

\subsection{Experimental Setup}
\noindent\textbf{Training Recipe:} For our experiments, we use a pre-trained 600M parameter Conformer encoder model~\cite{gulati2020conformer} with group~\cite{wu2018group} normalization (instead of batch normalization~\cite{ioffe2015batch} for better DP guarantees), fine-tuned with a CTC decoder~\cite{graves2006connectionist} and PEFT parameters on LibriSpeech~\cite{panayotov2015librispeech} (CC-BY-4.0) for 100k steps.
We implement our experiments in PAX~\cite{pax} and run them on 8$\times$8 TPUs v3.

\noindent\textbf{Hyperparameters:}
Across all DP experiments, we set the clipping bound to $2.5$ and fix $\delta=3.52\mathrm{e}{-6}$, smaller than the inverse of the training data size, as is standard.
We set the noise multiplier to achieve $(10.0, 3.52\mathrm{e}{-6})$-DP.
Note that according to recent work~\cite{ponomareva2023dp}, such a level of DP can be classified in the ``Tier 2: Reasonable privacy guarantees".  
We perform grid searches over other hyperparameters including learning rates and PEFT-specific hyperparameters such as the projection rank in LoRA.
Due to space limitations, we only present the best results unless otherwise stated.

{\bf \noindent Evaluation Metrics:}
Our main results include WER on LibriSpeech's two test-sets: test-clean and test-other, denoted as WER\_clean and WER\_other.
Due to space limit, ablation studies only report WER\_other, as it's more representative 
and performance trends align across both test splits.

\subsection{Adapting PEFT to ASR}
\label{subsec:adapt_peft_to_asr}
Most PEFTs originate from the language domain, and thus are mainly tested on language models.
To adapt them to ASR, we conduct various ablation studies as detailed below.

{\bf \noindent Choice of Bias Terms in BitFit:}
We observe that training all bias terms in the ASR encoder leads to divergence in BitFit.
Through a neural architecture search, we discover that freezing bias terms in layer normalization resolves this issue, and we adopt this approach for all BitFit experiments.

\begin{figure}[!t]
    \centering
    \subfloat[Non-private setting]{
        \includegraphics[width=0.48\textwidth]{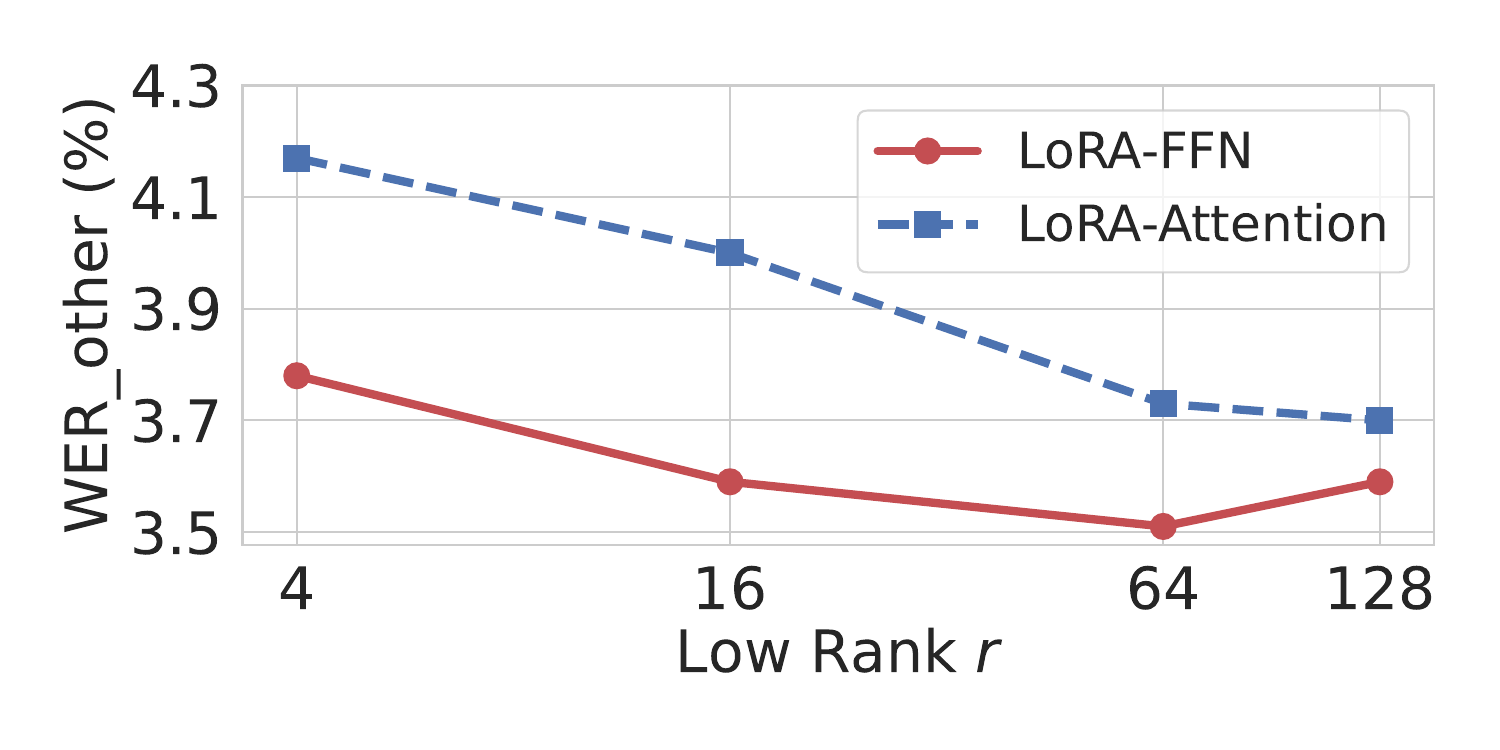}
        \label{fig:lora_wer_non_private}
    }
    \subfloat[$(10, 3.52\mathrm{e}{-6})$-DP]{
        \includegraphics[width=0.48\textwidth]{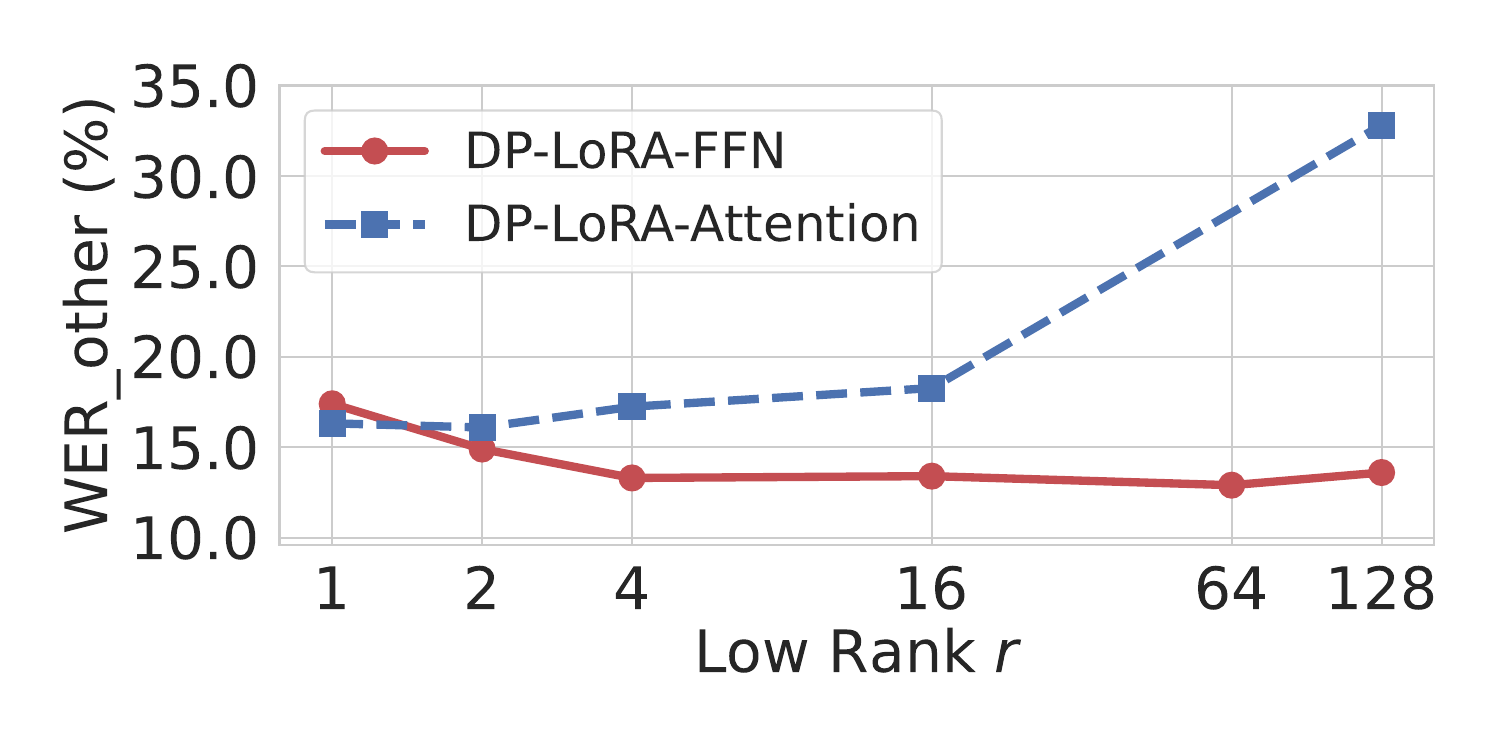} 
        \label{fig:lora_wer_dp}
    }
    \caption{WER\_other for different placements of LoRA modules.}
    \label{fig:lora_wer}
\end{figure}

{\bf \noindent Placement of LoRA Modules:}
We add LoRA to different Conformer layers, either in feed-forward (FFN) or self-attention layers, and compare their performance.
As shown in Figure~\ref{fig:lora_wer_non_private} and~\ref{fig:lora_wer_dp}, LoRA-FFN consistently outperforms LoRA-Attention, with or without DP, contrasting with the common practice of applying LoRA to self-attention in Transformers~\cite{hu2022lora}.

We also experiment with freezing the downscale matrix in LoRA, drawing inspiration from the concept of random projections~\cite{bingham2001random}.
Despite halving the trainable parameters LoRA, RP demonstrates a remarkably competitive performance, with only a minor drop in results, as detailed in Section~\ref{subsec:same_batch}.

\begin{figure}[!t]
    \centering
    \includegraphics[width=0.5\textwidth]{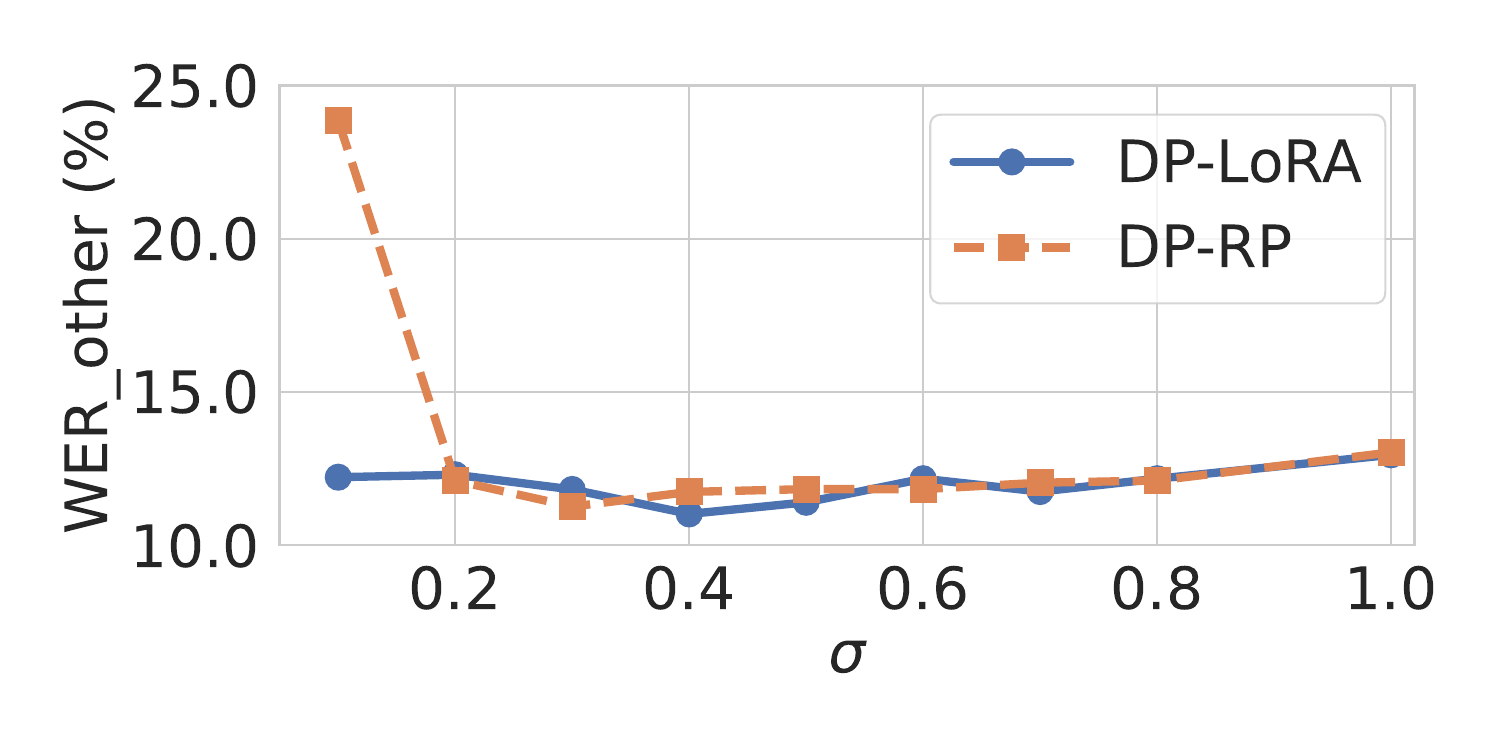}
    \caption{Impact of $\sigma$ in Gaussian initialization on DP-LoRA/DP-RP with $(10, 3.52\mathrm{e}{-6})$-DP.}
    \label{fig:impact_sigma}
\end{figure}

{\bf \noindent Initialization in DP-LoRA and DP-RP:} Following~\cite{hu2022lora}, we use a zero-centered Gaussian initialization with standard deviation $\sigma$ for downscale projection matrices in DP-LoRA and DP-RP.
We find that WER\_other initially decreases then increases with increasing $\sigma$ for both methods (Figure~\ref{fig:impact_sigma}).
The optimal $\sigma$ values for DP-LoRA (0.4) and DP-RP (0.3) result in WER\_other of 11.0\% and 11.3\%, respectively, highlighting the importance of proper initialization for these methods.
This key observation inspires us to leverage low-quality synthetic audio data to find a better initialization as described in Section~\ref{subsec:continued_optimization}.

\begin{figure}[!t]
    \centering
    \subfloat[Non-private setting]{
        \includegraphics[width=0.4\textwidth]{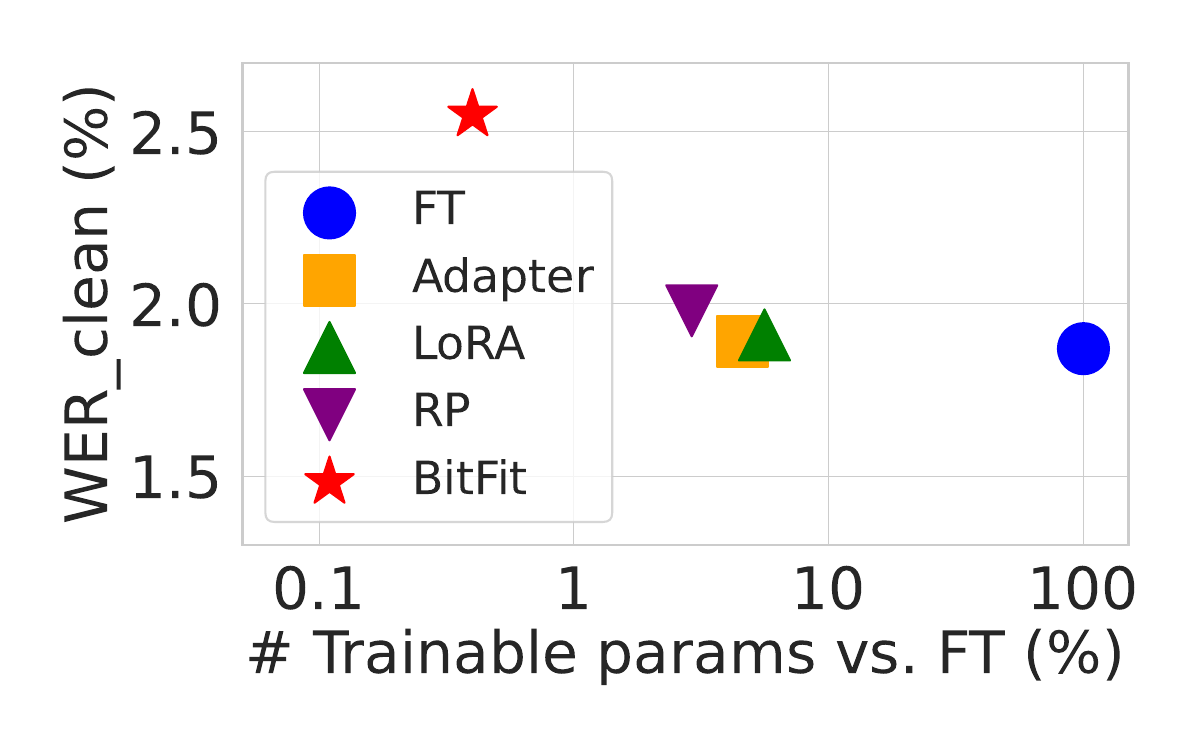} 
        \includegraphics[width=0.4\textwidth]{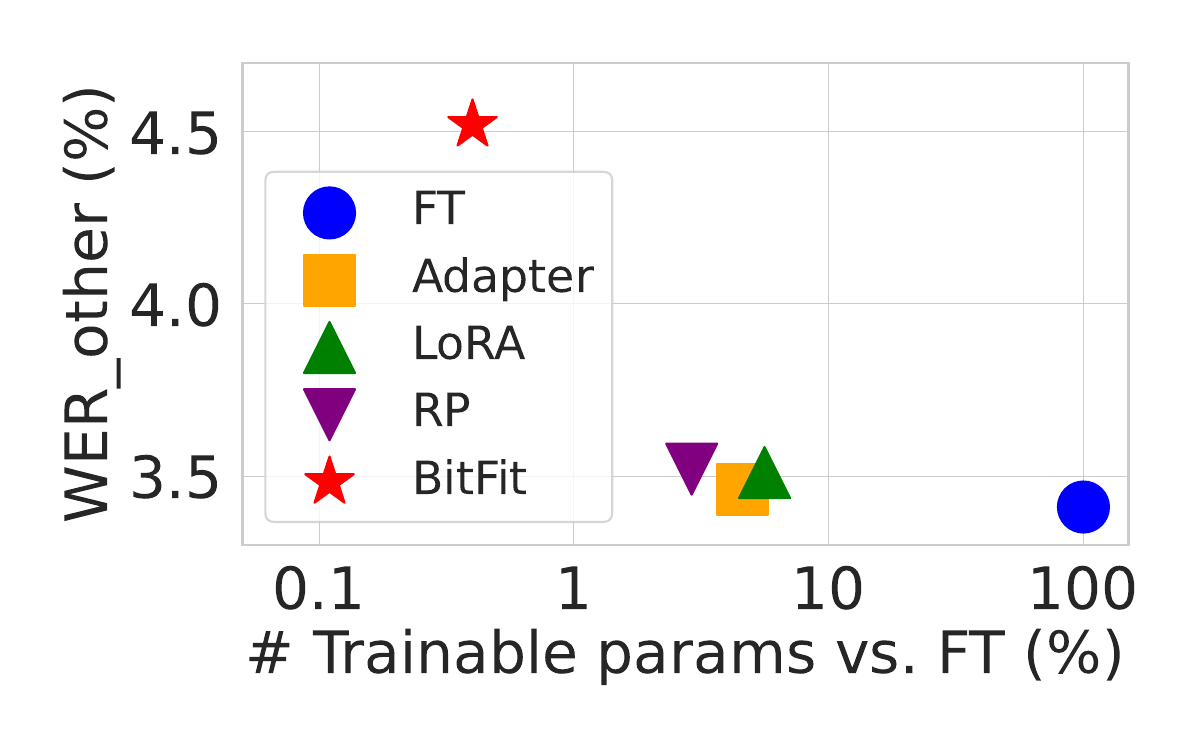}
        \label{fig:comparison_ft_vs_peft_non_dp}
    }
    \\
    \subfloat[$(10, 3.52\mathrm{e}{-6})$-DP]{
        \includegraphics[width=0.4\textwidth]{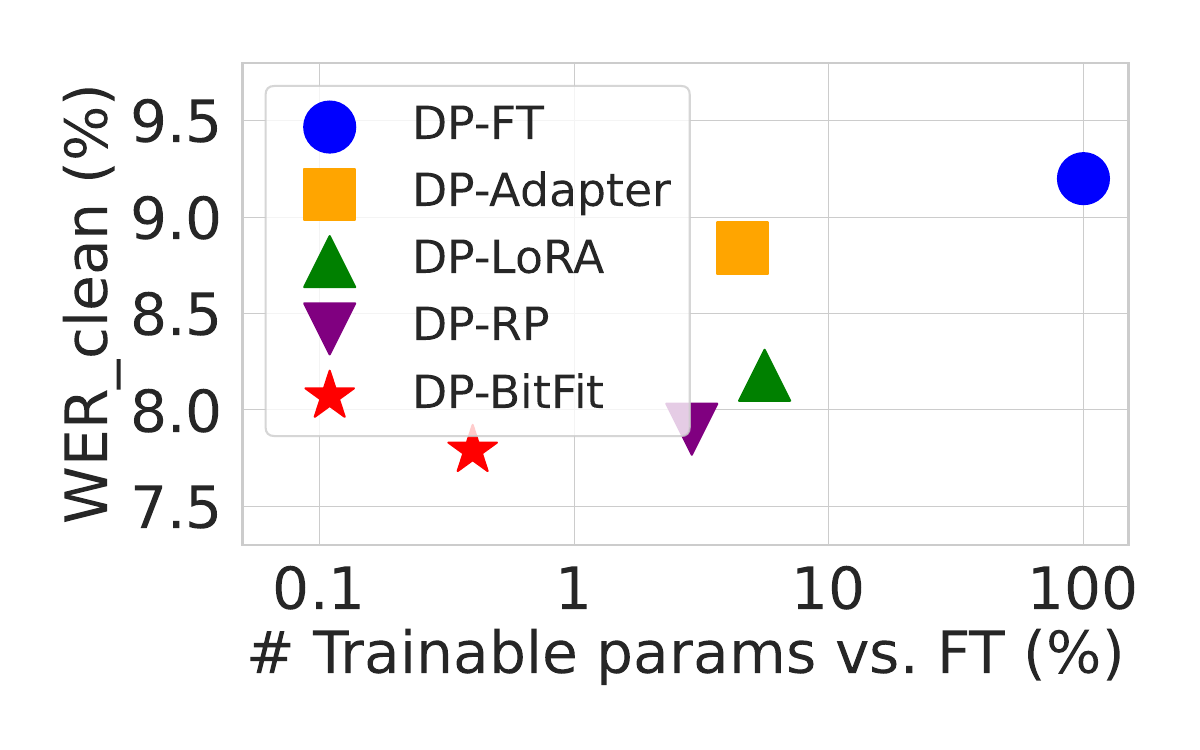} 
        \includegraphics[width=0.4\textwidth]{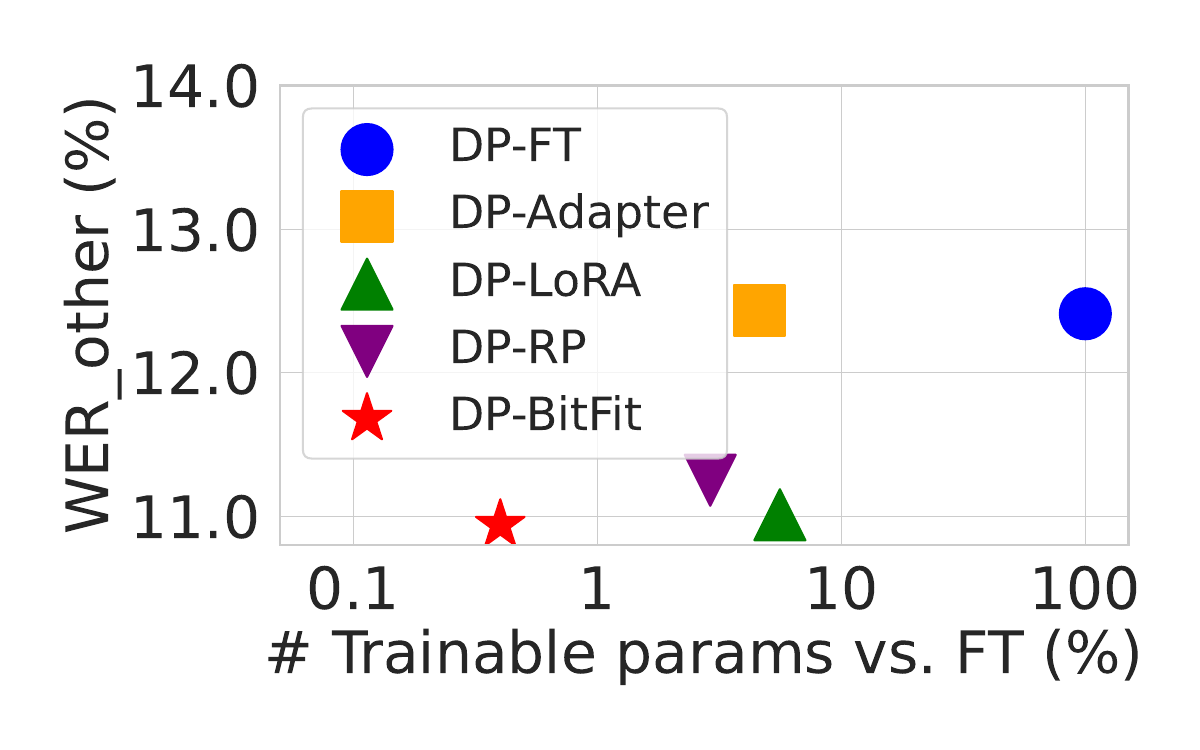} 
        \label{fig:comparison_ft_vs_peft_dp}
    }
    \caption{Comparison results of WER\_clean and WER\_other for DP FT/PEFT methods for the same number of fine-tuning steps.}
    \label{fig:comparison_ft_vs_peft}
\end{figure}

\subsection{PEFT (without DP) for ASR}
As a baseline, we compared PEFT methods without privacy guarantees and the results are summarized in Figure~\ref{fig:comparison_ft_vs_peft_non_dp}.
First, full fine-tuning achieves the best WER (1.9\% clean, 3.4\% other).
Among PEFT methods, Adapter and LoRA perform the best (1.9\% clean, 3.5\% other).
BitFit has the largest performance drop despite being most parameter-efficient.
Remarkably, RP achieves near-LoRA performance despite using only half the trainable parameters, establishing it as a compelling alternative for ASR applications with limited memory resources.

\subsection{DP-PEFT for ASR: Same Batch Size}
\label{subsec:same_batch}
Next, we compare WERs between DP-FT and DP-PEFTs at $(10, 3.52\mathrm{e}{-6})$-DP, as summarized in Figure \ref{fig:comparison_ft_vs_peft_dp}.
Surprisingly, DP-BitFit emerges as the optimal method in both parameter efficiency and WER.
It achieves the best WERs (7.8\% clean, 10.9\% other) by only fine-tuning 0.4\% of parameters compared to FT.
This phenomenon likely stems from ASR models prefer a lower-dimensional space and the landscape outside the low-dimensional space is curvy.

\subsection{DP-PEFT for ASR: Same Compute}
\label{subsec:same_compute}
Prior work~\cite{mcmahan2017learning,kairouz2021dpftrl,de2022unlocking} suggests that larger batch sizes can benefit private training in language and vision.
PEFT methods improve memory efficiency, so they unlock the opportunity to use larger batch sizes under the same compute constraints as DP-FT.
We conduct an ablation study by increasing the default 512 batch size by factors of 2, 4, 8, and 12, while adjusting training steps and noise multipliers to keep around 31 TPU-hours and $(10, 3.52\mathrm{e}{-6})$-DP.

\begin{table}[!ht]
  \centering
  \fontsize{10}{12}\selectfont
  \caption{Comparison results for DP FT/PEFT methods for the same TPU-hours under $(10, 3.52\mathrm{e}{-6})$-DP. \textbf{Bold} highlights optimal results.}
  \label{tab:comparison_dp_same_compute}
  \centering
  \begin{tabular}{c|c|c|c}
    \toprule
    \multirow{2}{*}{\makecell{Method}} & \multicolumn{2}{c|}{WER} & \multirow{2}{*}{\makecell{Optimal Batch\\Size Multiplier}} \\
    & clean & other &\\
    \midrule
    DP FT      & 9.2 & 12.4 &  1  \\
    DP Adapter & \textbf{6.0} & 9.2  &  8  \\
    DP LoRA    & 7.0 & 9.9  &  12 \\
    DP RP      & 8.5 & 11.3 &  2 \\  
    DP BitFit  & 6.1 & \textbf{9.1}  &  4  \\  
    \bottomrule
  \end{tabular}
\end{table}

Table~\ref{tab:comparison_dp_same_compute} compares DP-FT and DP-PEFT methods, showing all DP-PEFTs achieve lower WER than DP-FT when using similar TPU hours.
The performance ranking is basically consistent with that in Section~\ref{subsec:same_batch}, except for DP-Adapter, which improves significantly.
DP-BitFit still delivers the best WER\_other of 9.1\%.

\subsection{Using Low-quality Synthetic Data to Improve DP-BitFit}
\label{subsec:continued_optimization}

Considering the significance of initialization for DP-PEFTs highlighted in Section~\ref{subsec:adapt_peft_to_asr}, we explore the potential of using only low-quality synthetic audio data, consisting of random word transcripts, to achieve better initialization and enhance the performance of DP-BitFit.
This approach avoids privacy concerns associated with potentially sensitive information in transcripts and bypasses the difficulty of acquiring high-quality public datasets.

{\bf \noindent Synthetic Audio Generation:}
We generate 20,000 synthetic utterances, each consisting of 7 randomly sampled words from the 10,000 most frequent words in LibriSpeech test-other.
These transcripts are then passed through a TTS pipeline~\cite{oord2018parallel} with four diverse voices  (2 male, 2 female) to create the synthetic data.
We then pre-trained both bias terms in the ASR encoder and a randomly initialized decoder on this synthetic data for 3,000 steps before applying DP-BitFit on LibriSpeech.

{\bf \noindent Evaluation Results:}
Table~\ref{tab:comparison_synthetic_data} shows the impact of synthetic data on DP-BitFit.
Leveraging synthetic data, the WERs are further improved to 4.6/8.1 for training using the optimal batch size multiplier found in Section~\ref{subsec:same_compute}.
This establishes a new performance benchmark, achieving the best publicly reported model quality for ASR models under $(10, 3.52\mathrm{e}{-6})$-DP to date.

\begin{table}[!t]
    \caption{DP BitFit with synthetic data pre-training and 512$\times$4 batch size under privacy budget $(10, 3.52\mathrm{e}{-6})$.}
    \label{tab:comparison_synthetic_data}
    \centering
    \fontsize{10}{12}\selectfont
    \begin{tabular}{c|c|c}
    \toprule
    test splits & WER\_clean & WER\_other \\
    \midrule
    w/o synthetic data & 6.1 & 9.1 \\
    w/ synthetic & \textbf{4.6} & \textbf{8.1} \\
    \bottomrule
    \end{tabular}
\end{table}

\section{Conclusion}

This work extensively studies different methods for privately fine-tuning large ASR models. 
Our results on a state-of-the-art 600M Conformer-based model show DP-BitFit's superior memory efficiency and model quality.
We further introduce a novel synthetic data approach to improve privacy-utility trade-offs.
Finally we achieve state-of-the-art WERs (4.6/8.1) on LibriSpeech test-sets with small DP budgets. 

\section{Limitations}

A limitation of this work is the performance gap between differentially private models and their non-private counterparts.
Despite achieving state-of-the-art results under DP constraints, the word error rates of our models remain notably higher than those trained without privacy considerations.
This underscores the inherent trade-off between privacy and utility in differentially private machine learning, a challenge that warrants further investigation to minimize performance loss while maintaining strict privacy guarantees for large ASR models.
\bibliographystyle{unsrt}
\bibliography{ref}

\end{document}